\documentclass[letterpaper,12pt]{article}   %% LaTeX 2e (preferred)
\usepackage{osajnl2} %% do not use with REVTeX4
\usepackage[draft]{hyperref} %% optional
\usepackage{rotating}  % to rotate my tables
\begin{document}

\title{The Hertz/VPM polarimeter:  Design and first light observations}
\author{Megan Krejny$^{1,*}$, David Chuss$^{2}$, Christian Drouet d'Aubigny$^{3}$, Dathon Golish$^{3}$, Martin Houde$^{4}$, Howard Hui$^{2,5}$, Craig Kulesa$^{3}$, Robert F. Loewenstein$^{6}$, S. Harvey Moseley$^{2}$, Giles Novak$^{1}$, George Voellmer$^{2}$, Chris Walker$^{3}$, and Ed Wollack$^{2}$}
\address{$^1$Northwestern University, Department of Physics and Astronomy, \\ 2131 Tech Drive, Evanston, IL, 60208, USA}
\address{$^2$NASA Goddard Space Flight Center, \\ Greenbelt, MD, 20771, USA}
\address{$^3$University of Arizona, Steward Observatory, \\ 933 North Cherry Ave., Tucson, AZ, 85721, USA}
\address{$^4$The University of Western Ontario, Department of Physics and Astronomy, \\  London, Ontario, Canada N6A 3K7}
\address{$^{5}$Oregon State University, Department of Physics, 301 Weniger Hall, Corvallis, OR 97331}
\address{$^6$University of Chicago, Yerkes Observatory, \\ 373 West Geneva St., Williams Bay, WI, 53191}
\address{$^*$Corresponding author: m-krejny@northwestern.edu}

\begin{abstract}
We present first results of Hertz/VPM, the first submillimeter polarimeter employing the dual Variable-delay Polarization Modulator (dual-VPM).  This device differs from previously used polarization modulators in that it operates in translation rather than mechanical rotation.  We discuss the basic theory behind this device, and its potential advantages over the commonly used half wave plate (HWP).  The dual-VPM was tested both at the Submillimeter Telescope Observatory (SMTO) and in the lab.  In each case we present a detailed description of the setup.  We discovered nonideal behavior in the system.   This is at least in part due to properties of the VPM wire grids (diameter, spacing) employed in our experiment.   Despite this, we found that the dual-VPM system is robust, operating with high efficiency and low instrumental polarization.  This device is well suited for air and space-borne applications.

\end{abstract}

\section{Introduction}\label{sec:intro}

It has been almost 60 years since the discovery of the polarization of starlight by magnetically aligned dust grains \cite{hiltner49,hall49,chandraxx}.  Since then, astronomical polarimetry has become a valuable and well-established tool to study a wide variety of astrophysical sources, from nearby star-forming regions to the radiation linked to the formation of our universe.  Polarized submillimeter and FIR light from thermally emitting aligned dust grains, both in the interstellar medium and around stars, allows us to map the plane-of-sky magnetic field in these regions \cite{lazarian03,cudlip82,hildebrand84}.  In star-forming regions, light along the line of sight is emitted by dust grains at different temperatures; to separate cooler dust contributions from the warmer dust, e.g., near protostars, observations at multiple wavelengths are required.  Astronomical polarimetry is also of great interest in cosmology.  Measuring the polarization of the CMB could provide an opportunity to study the universe during its first $\sim 10^{-32}$ seconds after the Big Bang, when the universe is thought to have gone through an inflationary epoch at an energy scale $\sim$12 orders of magnitude above those accessible to terrestrial particle accelerators. 

Whether observing star-forming regions or remnants of the early universe one faces the same primary challenge:  measuring polarized fluxes that are 10$^{-6}$ or less of the total incident flux.  For dust and gas clouds, polarizations are often of order 10$^{-2}$ of the total source flux, but very small (10$^{-6}$) compared to the atmospheric flux (for ground-based observations).  For the CMB, the polarized flux is a million times below the total CMB power.  The polarized signature from inflationary physics is expected to be even lower (10$^{-7}$ to 10$^{-9}$).  

Measuring a small polarized signal in the presence of a large, unpolarized background is challenging.  Noise from the background as well as time variations in the instrument and observing environment dominate the signal. Polarization modulation allows for encoding of the polarization signal, enabling a subsequent extraction of the signal from the more random data stream.

In this paper, we describe the implementation of a novel polarization modulator,  the dual Variable-delay Polarization Modulator, or dual-VPM.  The dual-VPM operates in reflection instead of transmission, and fully modulates the linear polarization state using only small translational motions.  These properties make the dual-VPM an attractive alternative to the conventional birefringent half wave plate (HWP) modulator for certain applications.   We review the basic theory behind the dual-VPM and describe the development and characterization of this device.  

We used Hertz, a polarimeter previously used at the Caltech Submillimeter Observatory (CSO), as a dual-polarization detector for our dual-VPM system.  This new Hertz/VPM polarimeter was tested at 350\,$\mu$m at the Submillimeter Telescope Observatory (SMTO), where we characterized the performance of the VPMs.  Follow-up data were later collected at Northwestern University.  We present the results from these two tests.  We find that the dual-VPM system is robust, operating with high efficiency and low instrumental polarization.

\section{Polarimetric Techniques}\label{sec:theory}

\subsection{Stokes Parameters}\label{ssec:sp}

Full characterization of electromagnetic radiation requires knowledge of its amplitude and phase.   For partially-coherent radiation fields, this information is encoded in the time-averaged correlations between orthogonal fields. These are well-parameterized by the \textit{Stokes Parameters}, which describe the total flux ($I$), linearly polarized flux ($Q$ and $U$), and circularly polarized flux ($V$).  These four quantities are related by:

\begin{equation}\label{eq:iquv}
I^{2} \geq Q^{2} + U^{2} + V^{2},
\end{equation}
\\
where equality holds for fully polarized light.  Following the convention of \cite{jackson99}, we denote (for light propagating towards the viewer) Stokes $Q$ to be the difference between horizontal and vertical polarization, and Stokes $U$ as the difference between linear polarization oriented +45$^{\circ}$ and -45$^{\circ}$.  (The angle of polarization is defined to increase counterclockwise from the horizontal.)

\subsection{The Half-Wave Plate and the Variable-delay Polarization Modulator}\label{ssec:hwpvpm}

The half-wave plate (HWP) is a device that is able to induce a half-wave phase delay between incident orthogonal polarization components.  For this paper (unless otherwise specified) we will use the term ``HWP'' to mean a birefringent device  (for example, a quartz crystal) cut such that its optic axis is parallel to the front and back surfaces (orthogonal to the direction of light propagation), and whose thickness is chosen such that the plate will induce a 180$^{\circ}$ phase delay between orthogonal polarization components.  A rotation of the HWP by an angle $\theta$ causes a rotation of the plane of polarization by a corresponding $2\theta$.  The HWP can be summarized as a device that induces a fixed delay between orthogonal components, in effect rotating the polarization basis.  

\begin{figure}[htbp]
   \centering
   \includegraphics[width=10.6cm]{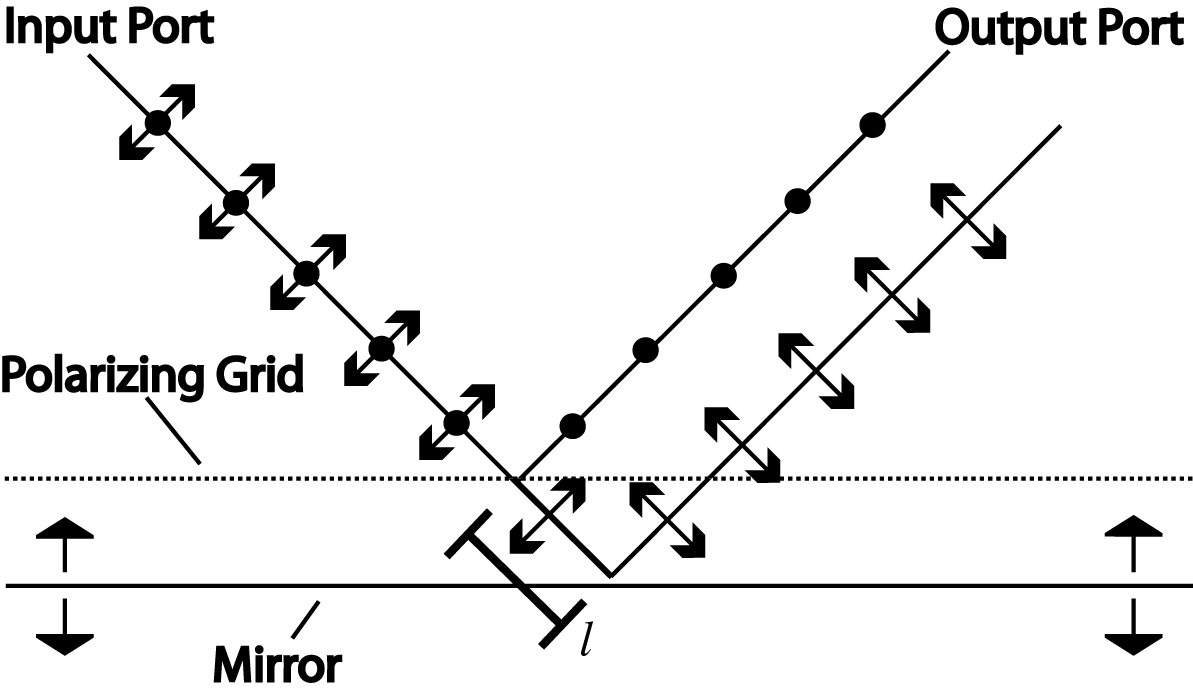}
   \caption{A ``reflective half-wave plate''.  Light incident upon a wire grid is separated into orthogonal polarization components; the component parallel to the wire grid is reflected, while the perpendicular component is transmitted and reflected by a mirror, traveling an extra distance $l$.  The case shown is for a 45$^{\circ}$ angle of incidence.  When the delay $l$ is set equal to half of the wavelength, this device has the same functionality as a birefringent half-wave plate (HWP)}
   \label{fig:concept}
\end{figure}

A  ``reflective HWP'' can be constructed by rotating a polarizing grid in front of a mirror.  Light incident upon the grid is separated into polarization either parallel or perpendicular to the grid wires.  The former is reflected by the grid, while the latter is transmitted to the mirror, reflects and then recombines with the orthogonal counterpart (Fig. \ref{fig:concept}).  For light with an incident angle of $\theta _{inc}$,  the induced path length difference $l$ is given by:

\begin{equation}\label{eq:vpmdisp}
l = 2d\cos \theta _{inc}   
\end{equation}
\\
If we set this path length difference to \textit{$\lambda/2$}, the device functions as a HWP in reflection.  Rotation of the grid is physically equivalent to rotating a HWP.  This device has been used for astronomical polarimetry at millimeter wavelengths \cite{shinnaga99, siringo04}.

Suppose that instead of maintaining a fixed distance between the grid and mirror surfaces, we now  move the mirror back and forth, thus changing the physical path length difference between the orthogonal polarization components.  By doing so, we fix the polarization basis and vary the phase delay; devices of this type have been denoted Variable-delay Polarization Modulators, or VPMs \cite{chuss06}.  

One VPM switched between half and full wave delays is equivalent to turning a HWP ``on'' or ``off'', that is, moving a HWP in and out of the beam.  We can calculate the necessary grid-mirror separation distances from Equation \ref{eq:vpmdisp}; for example, 350\,$\mu$m light incident at 20$^{\circ}$ requires settings of 93 and 186\,$\mu$m separations, respectively.  However, if polarization incident on the VPM grids is either completely parallel or perpendicular to the grid, then the polarization will not be modulated, regardless of the separation distance.  

\begin{figure}[htbp]
   \centering
   \includegraphics[width=14.0cm]{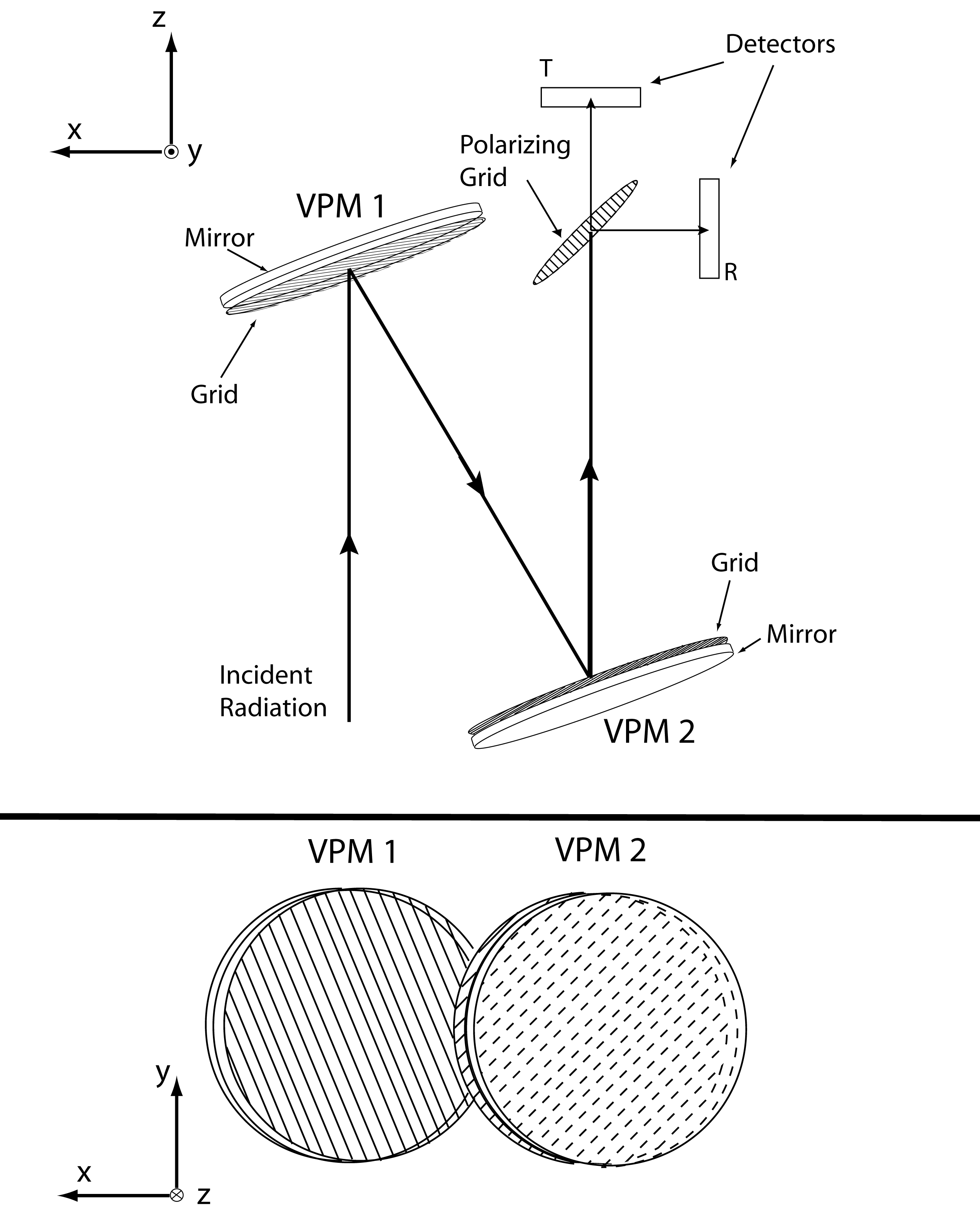}
   \caption{Two views of a schematic optical path for a polarimeter incorporating a dual-VPM modulator.  The upper panel shows a top view while the lower panel shows the view ``seen'' by the incoming radiation.  The radiation is reflected by two VPMs having their grid wires rotated by 22.5$^{\circ}$ with respect to one another, and is then incident upon a polarizing grid that splits the beam into orthogonal polarization components directed to detectors $R$ and $T$ (standing for reflected and transmitted).}
   \label{fig:overview}
\end{figure}

This problem can be solved by placing two VPMs in series, with their grids rotated by 22.5$^{\circ}$ with respect to one another (45$^{\circ}$ in Stokes space; see figure \ref{fig:overview}).  In this way any polarization not modulated by one VPM will be modulated by the other.  The dual-VPM system can thus accurately reproduce the function of a rotating HWP (provided that the analyzer grid of the polarimeter in Figure \ref{fig:overview} is not aligned parallel or perpendicular to the grid wires of VPM\,2) \cite{chuss06}. 

Single VPMs have been used for astronomical polarimetry at millimeter wavelengths in the form of a modified Martin-Puplett Interferometer \cite{battistelli02}.  However, since only one VPM was used, another modulator (double-fresnel rhomb) had to be incorporated.  The dual-VPM modulation scheme has the advantage of requiring only small translational motions, rather than rotation, to obtain full modulation of all linear polarization states.  This paper reports the first astronomical observations using a submillimeter polarimeter incorporating a dual-VPM modulator.

The VPM has several advantages over the HWP.  The VPM, in contrast to the birefringent HWP, operates in reflection, and so avoids some of the drawbacks of dielectrics.  Also, assuming near perfect performance over a large range of wavelengths, the wavelength of operation for the VPM can be easily tuned.  In comparison, for multiwavelength operation the dielectric HWP requires multiple birefringent layers and complicated (and often costly) achromatic antireflective coatings.  The VPM operates without a rotation bearing; the small motion of the mirror can be accomplished via piezoelectric motors and flexure bearings.  Flexure bearings  operate without friction and are generally considered to be the most durable of all non-levitating bearings.  Finally, the freedom of a variable delay means that the VPM can also act as a quarter-wave plate.  For broadband continuum work, the magnitude of astronomical circular polarization is often expected to be negligible; thus, measuring circular polarization can allow a check of the systematics of an experiment. 

The VPM does have challenges in its construction, characterization and operation.   The desire for high tolerances (we chose general tolerances below 10\,$\mu$m) requires careful design.  It also makes the device susceptible to large systematic vibrations.  Also, error caused by non-parallel grid and mirror surfaces must be carefully minimized, requiring longer setup times than for the HWP.  High grid efficiencies require flat and finely spaced wire grids, which can be expensive to produce.   Finally, as we shall see in Section \ref{ssec:int}, VPM characterization is nontrivial.  But first, we discuss basic polarization measurement techniques for the HWP and dual-VPM modulators.

\subsection{Data Acquisition and Analysis}\label{ssec:daq}

We begin with a discussion of the methods used to derive polarization signals from a dual-polarization detector for a single modulator position.   Below we start with a summary of basic data analysis techniques for submillimeter/millimeter wavelengths from \cite{hildebrand00}.  

Removal of the proportionately large sky signal present in our observations requires fast switching of the telescope beam between two points:  the source itself and an off-source point.  The chopping frequency must be fast enough to overcome $1/f$ noise from the atmosphere (often ranging from 3-15\,Hz).  As the secondary is chopping between on and off source, data points are calculated by subtracting the voltage value of the ``right'' beam from that for the ``left'' beam.  

We also ``nod'' the telescope, switching the source between the two beams.  This nodding technique reduces effects caused by surface defects and temperature differentials across the primary mirror.  We observe the object in a ``left-right-right-left'' ($l-r-r-l$) pattern, which allows us to eliminate long-term linear progressions in the signal, for example, if the telescope mirror were slowly heating over time.  Each nod in this ``chop-nod'' cycle contains an equal number of demodulated chop values; an average over all chops are saved for each nod, $l1, l2, r1$ and $r2$.  

To observe dual polarization simultaneously, a polarizing grid is used to direct orthogonal linear polarization components into two detector arrays such that one array observes the reflected light and the other transmitted light.  The intensity for one pixel in each array (reflected $R$ or transmitted $T$)  over one chop-nod cycle is calculated as:

\begin{equation}\label{eq:lrrl}
R (or\mbox{ }T) = (l1 - r1 -r2 + l2)/4
\end{equation}
\\
where the four terms refer to the averaged nod values as described above ($r$ terms are subtracted, since they are negative values).  The measured polarization signal $S_{mod}$ for a given modulator position is:

\begin{equation}\label{eq:RT}
S_{mod}=\frac{(R_{mod}-fT_{mod})}{(R_{mod} + fT_{mod})}
\end{equation}
\\
where $f$ is the relative gain between the corresponding $R$ and $T$ pixels being measured ($(\sum R)/(\sum T)$), averaged over all chop values for a full modulator cycle.  Stokes parameters are then calculated from the $S_{mod}$ values for one modulator cycle, in a manner determined by the type of polarization modulation used.

We define one ``HWP cycle'' as a stepped rotation of the device through a set of angles spaced evenly over a 180$^{\circ}$ degree range.   From Equation \ref{eq:RT}, the polarization signal $S_{mod}$ is plotted as a function of HWP angle, and a sine curve is fit to the data such that the amplitude of the curve gives the degree of polarization $P$, and the phase of the curve gives the polarization angle $\phi$.  We follow a fitting procedure similar to that explained by \cite{platt91}, taking into account the arbitrary offset of the zero HWP relative to our experiment.   We also adjust the sign of the HWP angle so as to account for the number of reflections between the HWP and the point of measurement.

For a ``VPM cycle'', the VPM grid-mirror separations are switched between ``on'' and ``off'' positions in four different combinations (VPM\,1-VPM\,2):  pos1:  on-on, pos2:  on-off, pos3:  off-on and pos4: off-off. Here,  ``on'' refers to a half-wave delay and ``off'' refers to a full-wave delay.   Unlike the HWP, the resulting polarization signals do not correspond to a sine curve; instead, we calculate the Stokes parameters as follows \cite{chuss06}:

\begin{equation}\label{eq:q}
q = [(S_{pos2} - S_{pos1})/2
\end{equation}

\begin{equation}\label{eq:u}
u = [(S_{pos3} - S_{pos4})/2
\end{equation}

These equations are consistent with the convention that the time-reversed polarized light beams from the $R$ and $T$ detector arrays reach VPM\,2 with polarization angles rotated by $\pm45^{\circ}$ with respect to the grid wires of VPM\,2 (e.g., see Figure \ref{fig:overview}).  This condition is necessary for full modulation of the polarization signal.  For our experiment, $R$ is sensitive to Stokes $-U$ and $T$ is sensitive to $+U$.  If the arrays were flipped in sensitivity, we would simply set $q$ to $-q$ and $u$ to $-u$.   $P$ and $\phi$ then follow from Stokes definitions:

\begin{equation}\label{eq:P}
P = \sqrt{q^{2} + u^{2}}
\end{equation}

\begin{equation}\label{eq:phi}
\phi = (1/2)\arctan(u/q)
\end{equation}
\\
Here we define the angle $\phi$ relative to the coordinate system of our optics plane.  

These cycles were defined for a system with only one modulator installed. When both are present, as we shall see is the case with our Hertz/VPM polarimeter, the above definitions are still applicable, provided one modulator is held fixed while the other is cycled.

 \section{Design of Hertz/VPM}\label{sec:hvpm}

Having described the theory behind the function of both the HWP and dual-VPM polarimeter and their respective analysis techniques, we now describe the physical implementation of the Hertz/VPM polarimeter.  The Hertz/VPM polarimeter consists of the dual-VPM modulator and the CSO polarimeter, Hertz.  An optics train, including the dual-VPM modulator, was built and used in front of Hertz.  We tested the full Hertz/VPM system at (SMTO) on Mt. Graham in Arizona from April 16-24, 2006.  The following year, we conducted a series of tests to further characterize Hertz/VPM in the lab at Northwestern University.  Below we will outline the construction of the VPMs and the experiment carried out at both SMTO and the lab.  This includes a description of the optics train as well as our use of  the Hertz polarimeter.
 
\subsection{VPM Construction}\label{ssec:const}

Our VPMs have four main elements:  1) an aluminum frame, consisting of a rectangular box with the top and back panels removed, 2) an optical quality mirror on a translation stage, mounted inside to the bottom of the frame, 3) a wire grid mounted to the front of the frame, and 4) a piezoelectric actuator, mounted in front of the frame, that controls mirror motion \cite{voellmer06}.

\begin{figure}[htbp]
   \centering
   \includegraphics[width=14.0cm]{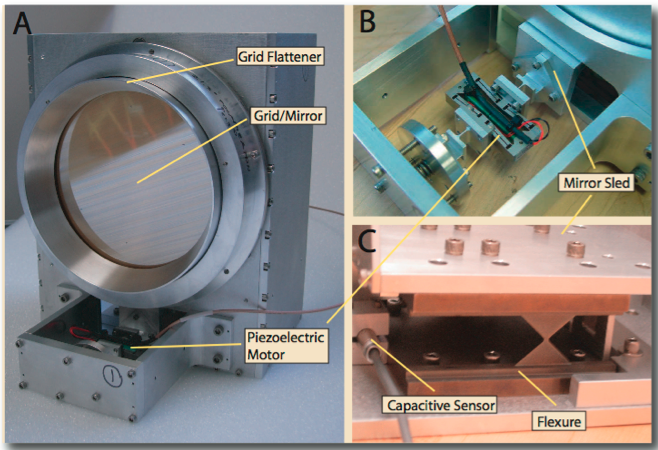} 
   \caption{Views of one VPM.  A:  Front view.  Wire grid is held to the front of the aluminum frame by rare earth magnets.  Grid flattener increases planarity of wires.  Its interior edge is milled to correspond with a 20$^{\circ}$ incident angle.  B:  View inside housing for piezoelectric actuator.  The motor is surrounded by titanium flexures which magnify the piezo motion.  The actuator is connected to two universal joint flexures to couple motion to the mirror and to absorb any twisting motion caused by misalignment.  C:  Back view under mirror mount, showing double-blade flexure and capacitive sensor.  Sensor measures actual mirror-grid separation distance and sends this information to servo controller, completing control loop.}
   \label{fig:VPM_views}
\end{figure}

The mirror is composed of vapor-deposited aluminum on glass and rests on a moving stage mounted on a kinematic variant of a double-blade flexure linear bearing.  The main flexures are machined in an hourglass shape so as to improve the parallelism of the bearing during motion.  This is shown in Figure \ref{fig:VPM_views}. The parallelism of the mirror motion was measured to be 1.5\,$\mu$m across the 150\,mm diameter mirror surface over a 400\,$\mu$m throw.   

Commercial piezoelectric actuators (DSM) control the mirror motion.  These motors rest perpendicularly to surrounding titanium flexures; elongation of the piezo pushes against these flexures, which subsequently magnify the motion (figure \ref{fig:VPM_views}). We measured the reliable full throw for the piezos to be 400\,$\mu$m.  Crossed flexure universal joints are coupled to each end of the piezoelectric motor.   The front universal joint is coupled to an adjustable plate on the front of the motor housing; this allows the user to define the actuator motion relative to the rectangular frame.  The back universal joint is connected to the moving mirror stage via an L bracket.  The universal joints act to prevent non-parallel motion of the actuator from affecting the mirror motion \cite{voellmer06}.  A servo control amplifier controls distance to within 1\,$\mu$m.  The grid-mirror displacement is measured using a capacitive sensor mounted to the bottom of the aluminum frame.

Grid quality was of primary concern in the construction of the VPMs, as error in wire spacing and grid flatness can create grid inefficiencies.   We used two 15\,cm diameter freestanding grids for the experiment.  Manufactured by Millitech, the grids consist of 25\,$\mu$m diameter gold coated wires, with a nominal spacing of 63\,$\mu$m.  We measured an actual average grid spacing of 67.5\,$\mu$m.   The grids are rated to frequencies up to 1600\,GHz (wavelengths down to 187.5\,$\mu$m), with a nominal efficiency of 95\% at this frequency.   

We desire an rms grid flatness that is less than 1\% of the operating wavelength.  Obtaining straight and parallel wires requires them to be under considerable tension.  This tension causes a ``potato chip'' effect, with the tension on the wires deforming the wire grid frame out of planarity.  Due to this effect, we measured the wires in the grids to have an rms flatness of roughly 35\,$\mu$m, equal to one tenth of our operating wavelength.  To improve grid efficiency, we developed a grid flattener.  The flattener has an optically flat end surface which rests against the stretched wire surface \cite{voellmer06}.  Set screws bring the flattener into contact with the wires just until the wires are deflected, minimizing the stress on the flattener itself.  The flattener was able to improve the rms flatness to $\sim$ 2\,$\mu$m.  Figure \ref{fig:VPM_views} shows the flattener; the interior surface was machined at a 20$^{\circ}$ angle in accordance with the beam incident angle.

The grid is pulled toward the VPM frame with rare earth magnets embedded in the front plate, while three set screws in the grid mount push against the frame, establishing the grid-mirror parallelism.  The set screws align with small divets located near the magnets on the front plate to establish repeatable rotational positioning of the wire grid.

\begin{figure}[htbp]
   \centering
   \includegraphics[width=14.0cm]{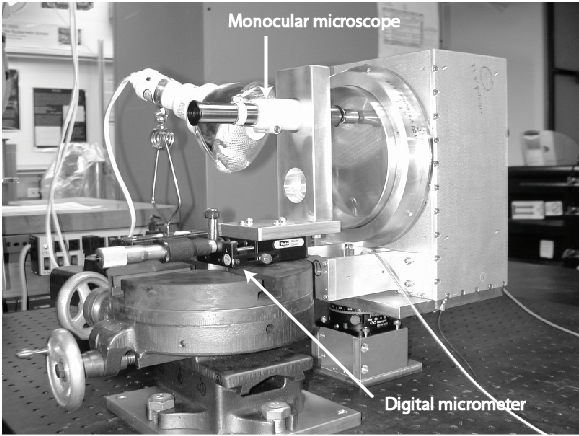}
   \caption{Microscope setup for grid-mirror parallelization measurements.  The microscope is mounted on a bracket that is moved toward and away from the grid via a linear translation stage that incorporates a digital micrometer.  Measurements are made at three points around the edge of the grid.}
   \label{fig:bench}
\end{figure}

Good parallelism between the mirror and grid surfaces is crucial to obtaining accurate phase delays and parallel polarization beams.  To set and measure the parallelism, we used a commercial monocular microscope with 200x magnification and mounted it on a linear translation stage with a micrometer, which in turn was mounted on a moveable base.  This setup is shown in Figure \ref{fig:bench}. The aluminum mount for the microscope has mounting holes at two different heights, corresponding to different measurement locations along the edge of the grid.  

To begin, we mount the grid to the front of the rectangular frame, aligning the set screws with the corresponding divets on the front surface.  Adjustments are made with the set screws to move the grid close to the mirror surface, moving one screw at a time.  A target separation distance is determined, calculated such that the mirror at its closest position is 50\,$\mu$m from the grid.  Using the microscope, the distance between the grid wires and their reflection is measured, and the set screws are adjusted until the separation distance equals the target distance. 

We repeat this procedure for each of three points on the wire grid; usually two or three iterations around the circle are needed.  The alignment process results in parallelism of roughly 5 $\pm$ 3\,$\mu$m.

\subsection{Optical Interface to Telescope}\label{ssec:expset}

Mt. Graham is located near Safford, AZ, and is operated by the University of Arizona.  The site rests at approximately 10,500 feet and has good submillimeter  ($\tau_{230\,GHz} \leq 0.06$) nights roughly 10-15\% of the time during the peak months of December-February, with a slightly lower percentage of good submillimeter nights in April.  We now describe the constructed optics train, the Hertz polarimeter, and the control system for the experiment.  

The Heinrich Hertz Telescope at the SMTO has a 10\,m primary and operates between 0.3 and 2\,mm wavelengths.  The focal ratio at the Nasmyth focus is 13.8 \cite{baars99}.  During our observations we chopped the secondary between two sky positions separated by 4$^{\prime}$ in cross-elevation at a rate of 3\,Hz.  

\begin{figure}[htbp]
   \centering
   \includegraphics[width=14.0cm]{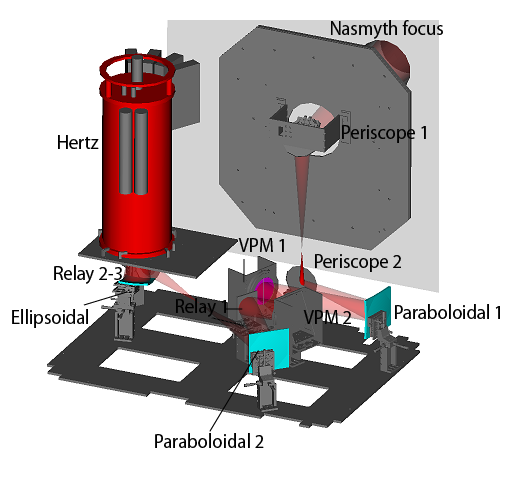}
   \caption{Optics path of Hertz/VPM experiment at SMTO.  Optics sit at Nasmyth focus of telescope.  Light is reflected down into the optics plane by two periscope mirrors.  Light is collimated with a paraboloidal mirror, then passes through the VPMs before being refocused by the second paraboloidal mirror.  After refocusing, a series of relay mirrors sends the light to an ellipsoidal mirror, which refocuses the light, this time to match the focal length of Hertz.}
   \label{fig:optics}
\end{figure}

Figure \ref{fig:optics} shows the optics path for the experiment.  Light from the telescope is incident upon two flat periscope mirrors that bend the light down into the horizontal optics plane.  The light is then collimated using an off-axis paraboloidal mirror before reaching the VPMs.  VPM\,1 has its grid wires rotated 22.5$^{\circ}$ counterclockwise from the horizontal, while the grid wires for VPM\,2 are aligned horizontally.  After passing through the modulators, the beam is then refocused using a second off-axis paraboloidal mirror.  Next, a series of two additional relay mirrors sends the beam to an ellipsoidal mirror which refocuses the light to match the focal ratio of the polarimeter (4.48).  Table \ref{tab:optics} lists optics components and their properties.   The optics were aligned in a two-step process; the first was a laser alignment of the optics carried out in the lab at Arizona.  Second, cold load tests were conducted at SMTO to ensure that the beam was centered on each optical element.  

In designing our dual-VPM system, one concern we faced regarding the optics was beam walkoff in the VPMs, the lateral translation between orthogonal components after they have passed through the mirror-grid system.  Simple geometry shows that the total walkoff can sum to a significant fraction of the wavelength.  We tried to minimize this effect in two ways.  The first was to place each VPM as close as possible to a pupil; lateral shifts at a pupil translate only to different incident angles at the focal plane.  This angular displacement is directly proportional to the incident angle at the VPMs.  Thus, to further minimize walkoff effects we made the incident angle as close to normal as possible.  We chose this angle to be 20$^\circ$, which allowed for sufficient beam clearance through the optics.  

%For some tests carried out at SMTO, a polarizing grid was mounted between the first paraboloidal mirror and VPM\,1, with its wires orientated horizontally.  Figure \ref{fig:mt} shows a photo of Hertz/VPM installed at SMTO.

\subsection{The Hertz Instrument}\label{ssec:hertz}

Hertz contains two 32-bolometer arrays (6x6 with the corners removed), cooled to 0.3\,K via a Helium-3 refrigerator.  An analyzer grid splits the incoming signal into two orthogonal linear polarizations and directs each to a detector array. This dual-polarization observing strategy results in a  $\sqrt{2}$ increase in the signal-to-noise over signal polarization systems, and also aids in removing sky noise, which is correlated between arrays \cite{hildebrand00}.   The polarimeter operates at 350\,$\mu$m, with a relative bandwidth of $(\Delta\lambda)/\lambda$ = 10\%. Hertz contains cold reimaging optics, using quartz lenses \cite{schleuning97}.

Hertz incorporates a quartz HWP, located at a cold pupil stop.  Although we did not require this for polarization measurements carried out with the dual-VPM polarimeter, we used it for two purposes:  first, by using the Hertz instrument in its original HWP-polarimeter mode, we were able to measure the linear polarization state at the output of the VPMs, providing a diagnostic of the dual-VPM modulator by itself.    Second, as mentioned in section \ref{ssec:daq},  polarization observations with Hertz/VPM require that the time-reversed polarized light beams from the $R$ and $T$ detector arrays reach VPM\,2 with polarization angles rotated by $\pm45^{\circ}$ with respect to the grid wires of VPM\,2 (e.g., see Figure \ref{fig:overview}).  It was convenient to achieve this condition by rotating the HWP in Hertz rather than having to rotate the entire dewar.  

We thus needed to determine the HWP angle that satisfied this criterion.  At the SMTO, we accomplished this by placing a polarizing grid with horizontal wires in the optics train, directly after VPM\,2.  We then rotated the HWP until the signals were approximately equal in the $R$ and $T$ arrays.  This angle was determined to correspond to an encoder reader of 96$^{\circ}$ (relative to an arbirary offset).  We collected data at both 96$^{\circ}$  and the equivalent angle of 51$^{\circ}$ (which switches only the sign convention of the measured Stokes parameter, as seen in section \ref{ssec:daq}).

\subsection{Control System}\label{ssec:control}

\begin{figure}[htbp]
   \centering
   \includegraphics[width=16.0cm]{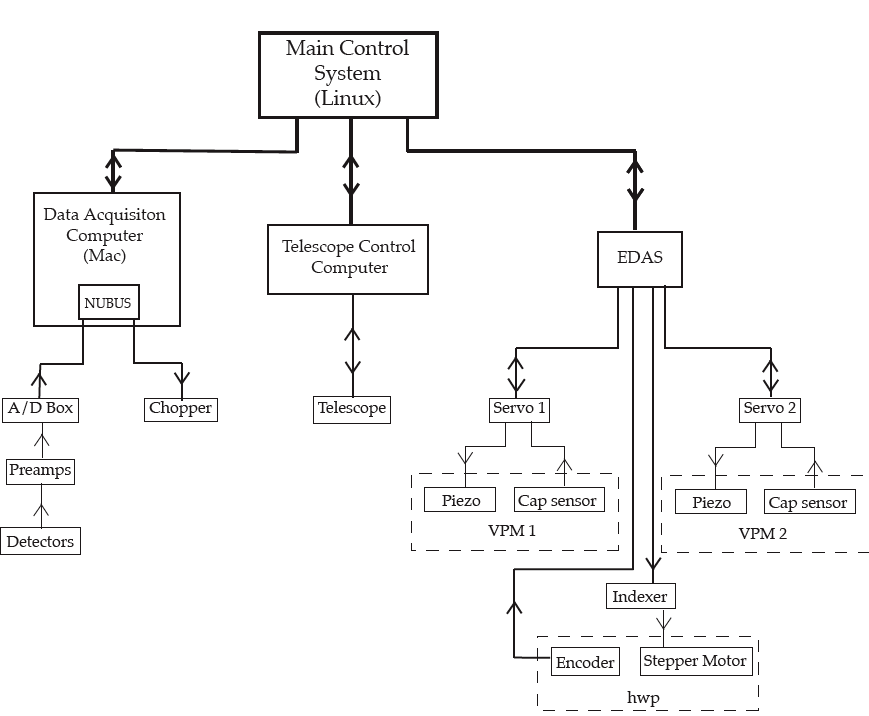}
   \caption{Diagram of control electronics.  Lines signify paths of communication between components; arrows signify direction.  Main control computer communicates with three computers via TCP/IP connections.  The data acquisition computer reads demodulated signals from detectors, synchronized by a chopping frequency sent to the secondary mirror.  Telescope control computer operates telescope motion and positioning.  EDAS controls both the VPMs and the Hertz HWP.}
   \label{fig:compcon}
\end{figure}

The control system is outlined in Figure \ref{fig:compcon}.  The main control computer sends commands to the other computers through a user-operated GUI.  The main control computer communicates via TCP/IP with three computers:  the telescope control computer, which controls the telescope motion and positioning, the data acquisition computer, which records bolometer output to file along with header information, displays data onscreen, and controls the chopping secondary mirror, and the Ethernet Data Acquisition System (EDAS) which operates the modulators (HWP and VPMs).

The data acquisition computer receives, stores and displays data sent from the Hertz detector.  Signals originate in Hertz as bolometer voltages that are amplified and then converted to a digital signal via an A/D converter.  The data acquisition computer incorporates a custom-built Data Signal Processor (DSP) card.    The computer records only the chopper-demodulated (i.e., on source -- off source) signals.  The DSP card performs the demodulation of the incoming data stream in synchronization with the chopper reference signal that it generates and sends to the secondary mirror controller.  After demodulation, the data acquisition computer saves the demodulated data for each pixel to disk. 

The EDAS (Intelligent Instruments) is a modular system computer that allows remote operation via an internet connection. Its use of flash memory eliminates the risk of hard drive failure.  Three modules are connected to the EDAS main CPU and power modules:  a serial port module, an analog output module (AO) and an analog input module (AI).  

To move the VPM mirrors, a target position is sent by the main control computer to the EDAS, which sends that command (via binary serial communications) to the servo controller.  The servo sends the actual position recorded by the capacitive sensor back to the EDAS (via binary serial) which relays it to the main control computer (via TCP/IP).  To rotate the HWP, the EDAS sends serial commands to a stepper motor indexer, which controls the HWP motor.  The EDAS measures the HWP positions by first applying an output voltage across a HWP encoder via the AO.  The custom-built HWP encoder consists of a cryogenic rotary variable resistor mounted to the HWP.  The output voltage across the resistor is read by the AI module and is converted to a HWP angle via a simple linear equation determined by previous measurements.  

\subsection{Laboratory Test Setup}\label{ssec:NU}

Next we describe the setup we used to carry out testing of the Hertz/VPM polarimeter at Northwestern University.  We simulated the telescope signal using a commercial blackbody source placed behind a rotating optical chopper.  The blackbody was placed near the focal length of the first paraboloidal mirror.  To facilitate the mounting and alignment of the blackbody, we removed the two periscope mirrors, replacing them with a flat mirror that kept the beam in the horizontal optics plane.  A polarizing grid was placed between the blackbody and the chopper as needed.  The grid was mounted on a rotating support, allowing us to vary the polarization state of the source.  We rotated the grid to different angles corresponding to $\pm Q$ and $\pm U$ polarization states.

Some difficulty arose due to extraneous signals caused by the bolometer array viewing an image of itself reflected by the aluminum aperture plate of the blackbody.  We were able to eliminate this effect by first rotating the blackbody, grid and chopper planes slightly from orthogonality to the beam, and then placing a new aperture, drilled from a piece of absorber, in front of the blackbody/grid/chopper system.  The optics alignment achieved at Northwestern was not as accurate as for the SMTO tests.  A careful laser alignment of the mirrors was carried out, but no cold load alignment was done.   This may in part explain the measured efficiency of the system.

At SMTO,  we determined the HWP angle setting empirically as described in section \ref{ssec:hertz}.   In the lab, we rotated the polarizing grid wires to an angle rotated 45$^{\circ}$ CCW from horizontal, i.e., $-U$ input polarization.  We took HWP files (each modulator file consists of one cycle) with VPM\,2 set to a full wave delay, and with the VPM\,1 grid removed.  We then looked at a plot of the polarization signal as a function of HWP offset angle.  The location of the first peak corresponded to the proper HWP angle.   We measured this angle to be 80$^{\circ}$, a difference of almost 16$^{\circ}$ from the previous setting (96$^{\circ}$).  Due to an error in the calculation, we also collected data at a HWP angle of 68$^{\circ}$.  Later, in section \ref{ssec:eff}, we will discuss how these HWP angle settings affected our measurements.
 
\subsection{Observations}\label{ssec:obs}

To fully characterize the Hertz/VPM polarimeter, we took HWP and VPM files with different input sources (SgrB2, planets, blackbody source),  either with or without a polarizing grid placed in the optics train.  Data taken without polarizing grids were used to obtain photometry maps and instrumental polarization (IP) measurements.  Data taken with a polarizing grid installed were used to characterize VPM performance and determine system polarization efficiency.  

We present SMTO data collected only on the last night of the run (April 24), as it was the only night of the run with low atmospheric opacity ($\tau_{230\,GHz} \leq$ 0.06) and stable observing conditions.   Data taken in the lab were collected over a two week period in October 2007.

\section{Results}\label{sec:results}

\subsection{Photometry of Sagittarius B2}\label{ssec:sgrb2}

\begin{figure}[htbp]
   \centering
   \includegraphics[width=12.0cm]{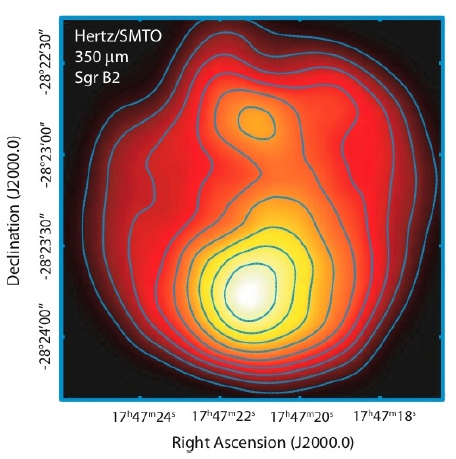}
   \caption{350\,$\mu$m photometric map of SgrB2, created from VPM files taken with the Hertz/dual-VPM polarimeter.  Files were added and smoothed with no flat fielding.  Two peaks are shown, $\sim$ 54$^{\prime\prime}$ apart, which agree with previous results at this wavelength \cite{goldsmith90}.  The beam size is 20$^{\prime\prime}$.}
   \label{fig:sgrb2}
\end{figure}

SgrB2 is a giant molecular cloud located 100\,pc from the Galactic Center.  Its dense cores are known to be centers of massive star formation; these regions have been mapped polarimetrically \cite{novak97}.  Figure \ref{fig:sgrb2} shows a smoothed photometric map of SgrB2 taken with the Hertz/VPM polarimeter.  Total integration time was 90 minutes (15 VPM files).  The files were combined, correcting for changing parallactic angle.  No flat-fielding was done.  The fluxes are normalized to unity at the peak.  

The map shows two peaks, separated by approximately 54$^{\prime\prime}$.  This agrees with previous observations \cite{goldsmith90}.  Based on the design of the system, we assumed the beam size to be similar to that of Hertz at the CSO (20$^{\prime\prime}$) \cite{dowell98}.  Given the relative faintness of the source, we consider our map to be evidence that our system has reasonable optical coupling.  

\subsection{VPM Interferograms and the Asymmetry Problem}\label{ssec:int}

At the SMTO, we observed Saturn through a polarizing grid mounted horizontally with respect to the optics plane.  We took HWP files with the VPMs set to various combinations of grid-mirror separation distances.  The data were fit using a 2-VPM transfer function model \cite{chuss06}.  This model allowed us to fit for the source polarization, small offsets in the grid-mirror separation of each VPM, and rotational errors in the alignment of the VPM grids. Though the data exhibited the same qualitative features of the model, the apparent phase delays of the VPMs did not match the delays one expects given the geometry, wavelength and reported grid-mirror separation. 

We isolated this problem in the lab by studying the polarization properties of a single VPM. Using the laboratory setup described earlier, we removed the grid from the front of VPM\,1 and sent polarized light oriented $-45^{\circ}$ with respect to the horizontal axis ($-U$) into VPM\,2, whose grid axis is aligned horizontal to the optics plane.  We then stepped the grid-mirror separation distance from 50 to 450\,$\mu$m, roughly two full wavelength cycles of polarization modulation, acquiring a HWP file at each position.

\begin{figure}[t!]
   \centering
   \includegraphics[width=14.0cm]{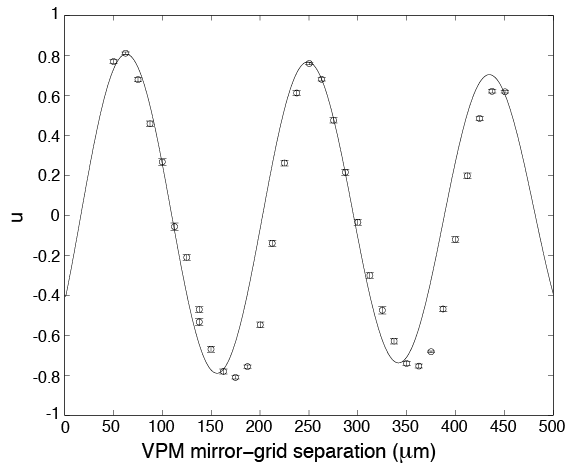}
   \caption{Interferogram plotting normalized Stokes parameter $u$ versus mirror-grid separation for VPM\,2.  Input light is assumed to be 100\% polarized at an angle of -45$^{\circ}$ with respect to the $+x$ axis.  The solid line is the signal expected for a geometric phase delay.  This ideal sinusoidal curve has been modified to match the amplitude and phase of the first peak of the observed data.  The curve also has a decoherence function incorporated to account for 10\% bandwidth effects of the Hertz polarimeter.  The data do not match this simple geometrically-motivated model; there is an asymmetry in the observed data, with the peak-valley separation distance less than the theoretical separation (93\,$\mu$m).  A more detailed model that includes the phase response of the grid is required to reproduce the observed instrumental performance.}
   \label{fig:interf}
\end{figure}

For ideal grid performance and monochromatic radiation, we expect a unity-amplitude sine modulation of $u$ with extrema located at the theoretical spacings (for $\lambda$ = 350\,$\mu$m) of (n+ 1/2)*186\,$\mu$m (VPM ``on'') and $n$*186\,$\mu$m (VPM ``off''), as we showed earlier in section \ref{ssec:hwpvpm}.  Figure \ref{fig:interf} plots normalized Stokes $u$ versus grid-mirror separation distance for VPM\,2; each point plotted represents a HWP file taken at one VPM setting.  The solid curve represents the theoretical performance of an ideal VPM, taking into account both the observed system efficiency and decoherence due to the finite bandwidth of the system (see section \ref{ssec:hertz}).  For illustrative purposes, we applied an arbitrary offset to align the curve with the first peak of the lab data.   We notice an asymmetry between the ascending and descending portions of the curve.  Although the separation between adjacent peaks (or adjacent valleys) is roughly 186\,$\mu$m, the peak-valley separations are smaller than predicted.  We also observe an efficiency well below 100\%; a  discussion of possible causes of the observed efficiency is deferred until section \ref{ssec:eff}.  

Models exist in the literature for wire grid performance.   We have used the results of \cite{houde01} to attempt to reproduce the aforementioned asymmetry observed in our experimental results for the normalized Stokes $u$ parameter as a function of the grid-mirror separation.  We ran a simulation of the corresponding theoretical response of our VPM.  The model used the wire radius $a$ and spacing $d$ of our experiment, and conductivity values $\sigma$ that corresponded to aluminum-coated glass mirrors and gold plated wires (see section \ref{ssec:const}).  

%The grids considered in the model were assumed to consist of an infinite number of wires of infinite length, with no error in grid wire spacing.

\begin{figure}[htbp]
\centering
\includegraphics[width=12.0cm]{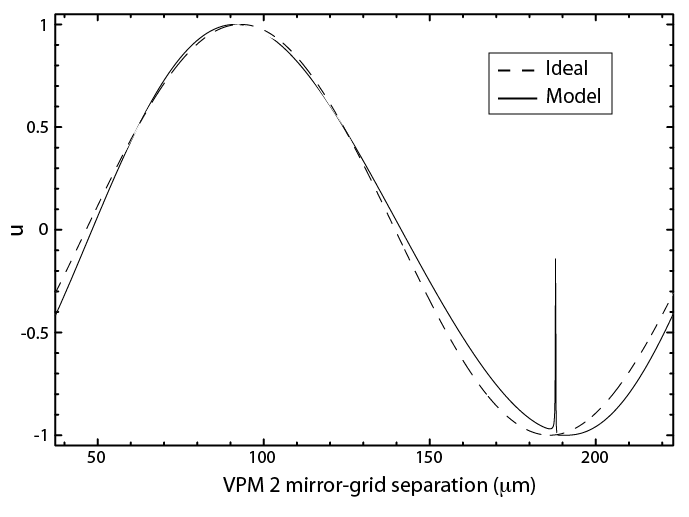}
\caption{Interferogram obtained from ``reflecting polarizer'' model \cite{houde01}.  Solid curve represents model predictions ($u$ versus grid-mirror separation distance) for grid parameters and incident polarization that match our experiment.  The dashed curve shows the case of perfect reflection and transmission of polarized light oriented parallel and orthogonal to the grid wires, respectively.  The model assumes infinite wire grids of infinite length with 100\% efficiency and no error in wire separation.   Spike at minimum is broadened for large relative bandwidth and is not seen with Hertz/VPM.   In the long wavelength limit, the model approaches the ideal curve.}
\label{fig:houde}
\end{figure}

The dashed line in Figure \ref{fig:houde} represents the behavior of an idealized grid that completely reflects light with polarization parallel to the grid and completely transmits light polarized perpendicular to the grid. Note that this plot is consistent with a pure sinusoid.   The solid line of Figure \ref{fig:houde} shows the simulated plot of Stokes $u$ versus grid-mirror separation distance for a narrow bandwidth ``reflecting polarizer.''  Similar to the lab data, the model shows an asymmetry in the data curve.  The difference between the ideal and predicted curves is a function of the grid parameters $a$ and $d$ and the operating wavelength, $\lambda$.  Increasing the wavelength by a factor of 10 in the model eliminates the observed asymmetry.  In our case, $\lambda/a= 350/12.5 = 28$.  This is somewhat outside of the strict applicability of the model \cite{houde01} ($\lambda/a>40$), which is the likely cause of the remaining discrepancy between the model and observed data.  (We can ignore the spike at the minimum, which is a feature of the low bandwidth of this model, and is broadened out for Hertz/VPM.)  However, the general trend is observed.

It is important to note that even though we find that the actual phase delay differs from the geometric prediction, the effect described here does not affect the utility of the VPM.  It simply indicates that a more detailed model is required to map the grid-mirror separation into phase.  For the purposes of this work, the interferogram in Figure \ref{fig:interf} can be used to ``tune'' our VPMs. Our interferogram shows that, for the VPM grids used, operating at 350\,$\mu$m, the peaks and valleys are not at the theoretical values of 93 and 186\,$\mu$m.  Presumably, adjustments to the mirror-grid separation distances so that the full- and half-wave conditions are met will increase the polarization efficiency of the system.  After locating the proper full- and half-wave delays for VPM\,2, we then set the device to a full-wave delay (i.e., turned ``off''), and repeated the interferogram measurements, this time slowly stepping VPM\,1.  We took data both at these observed separation distances and other settings related to the theoretical separations, which were later used for comparison.

\subsection{Efficiency measurements from grid tests}\label{ssec:eff}

As suggested by the interferogram plot in the previous section, there are sources of polarization inefficiency in the experiment.  We now discuss a series of grid measurements taken to isolate the causes of polarization efficiency losses.

Some sources can be found within Hertz; namely, its HWP and polarizing grid.  In the lab, we collected HWP files taken with a polarizing source grid and with both VPM grids removed.  We took files for four polarizing grid angles, corresponding to input polarization of $\pm q$ or $\pm u = -1$.   With this configuration, we measured an average efficiency of 93\%.  Previous measurements for the Hertz polarimeter give an efficiency of 95\% \cite{dowell98}; this drop of 2\% may be attributable to the quality of the source polarizing grid, which has not been measured.   

There are two main factors that contribute significantly to the efficiency measurements of the Hertz/dual-VPM system: the HWP offset angle setting and the VPM separation distance settings.  We compare SMTO and lab data taken under three different conditions:  incorrect HWP and VPM settings,  incorrect HWP setting and correct VPM settings, and correct HWP and VPM settings.  In total, there are five different groups of data; the parameters for these groups are given in Table \ref{tab:settings}.  There are three groups that fall under the first set of conditions listed above; two taken at SMTO and one in the lab.  The determination of the HWP offset angle settings were discussed previously in sections \ref{ssec:hertz} and \ref{ssec:NU}.  We note that for group 2, the HWP angle is closer to (80-45 =) 35$^{\circ}$ than our proper offset angle of 80$^{\circ}$; thus, our determination of Stokes $q$ and $u$ for this group is the opposite convention as explained in section \ref{ssec:daq}.  Also note that for group 4,  the difference between half and full wave delays was 93\,$\mu$m, but the grid-mirror separations were offset from the theoretical 93/186\,$\mu$m positions.   

For SMTO grid data, we took nine VPM files (four for group 1, five for group 2) of Saturn with a polarizing grid installed before the first paraboloidal mirror.  For each of these two groups, files were taken at one polarizing grid angle (horizontal, or $-Q$ input polarization).   For the lab data, group 3 had six files taken at each of four polarizing grid angles ($\pm Q, \pm U$), while groups 4 and 5 had two files taken at each angle.  

\begin{figure}[htbp]
   \centering
   \includegraphics[width=16.0cm]{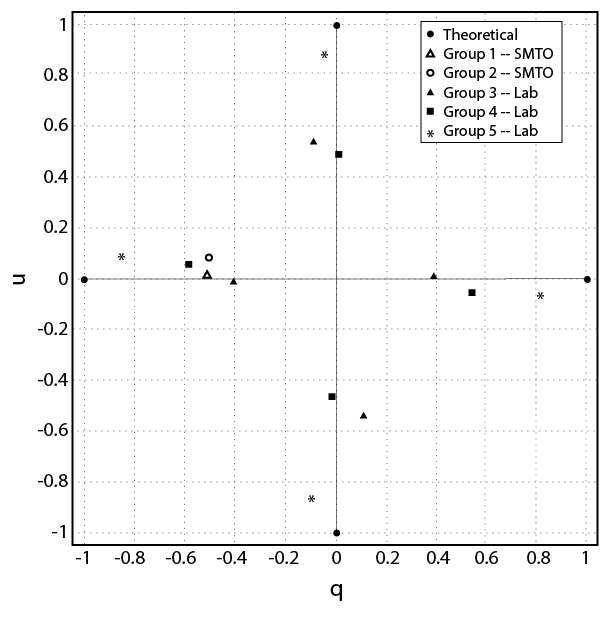}
   \caption{ Polarization efficiency of Hertz/VPM polarimeter.  Circles on axes represent 100\% input polarization.  Points plotted represent average values calculated for each group.  In $qu$ space, points are rotated away from axes; in real space, polarization angles are within 3$^{\circ}$ of theoretical values.  Overall efficiencies increase as HWP and VPMs are moved to their proper settings.}
   \label{fig:qu}
\end{figure}

For each VPM file we calculated normalized Stokes $q$ and $u$ by the method outlined in section \ref{ssec:daq}.  Figure \ref{fig:qu} shows the input polarizations and the average efficiency for each group.  In addition to plotting the efficiencies, we calculated the total efficiency for each group by averaging over all files in a group, regardless of whether that file was a measure of $q$ or $u$-like polarization.  Averages were calculated with equal weighting and are listed in Table \ref{tab:settings}.

Figure \ref{fig:qu} shows the efficiency measurements for incoming polarization at four different grid angles:  0$^{\circ}$ and 90$^{\circ}$ ($-Q$ and $+Q$ polarization), 45$^{\circ}$ and 135$^{\circ}$ ($-U$ and $+U$ polarization).  Points that lie on the unit circle represent 100\% input polarization, which we assume is the efficiency of the polarizing grid.  As can be seen, an incorrect HWP or VPM position affects the polarization efficiency, which means that we can accept our Mt. Graham results with only a multiplicative adjustment to compensate.  It is important to note, however, that there are variations in the polarization efficiency as one travels around the Stokes plane.  For example, $q$ polarization efficiency is higher than $u$ efficiency for groups 3 and 5, but the reverse is true for group 4.  This phenomenon is not fully understood; however, it is evident that using the proper modulator settings results in an overall increase in polarization efficiency.  The polarization efficiency ranges from 40\% to a maximum of roughly 85\% for group 5, where the HWP and VPMs are set to their optimal positions. Polarization efficiency at the SMTO (groups 1 and 2) was measured to be 54\%.  

Even when the VPMs are set to the proper half- and full-wave delays, we still see only a maxiumum efficiency of 85\%; for perfect efficiency, we should expect a polarization closer to 93\%.  We do not know the cause of this net dual-VPM efficiency loss.  Possible causes might include poor optical alignment in the lab or efficiency losses due to the polarizing grid, as mentioned above.  

The polarization points do not line up along the axes, but rather appear rotated slightly about the origin.  In physical space, all measured polarization angles are within 3$^{\circ}$ of the expected values.  We believe this error to be due to cumulative alignment errors.  

\subsection{Instrumental Polarization}\label{ssec:ip}

We observed Jupiter with no polarizing grid on the last night of observations at SMTO.   The planet is assumed to be unpolarized \cite{clemens90}; hence any detected polarization should be attributed to the instrumental polarization (IP) of the full Hertz/VPM system.  We collected 15 VPM files for group 1 and 14 files for group 2.

Similar to the polarized grid files, we calculated the normalized $q$ and $u$ values for each VPM file.  Averages were calculated for each group with equal weighting, and the $q_{ave}$ and $u_{ave}$ values were combined to give a percent polarization $P$ and angle $\phi$.  We calculated raw values for $q$, $u$ and $P$, then recalculated these values by a gain factor $g$ of $1$/(\textit{polarization efficiency}), with the polarization efficiency given by the values calculated in the previous section.

We measured raw IP values of 0.29 $\pm$ 0.06\% and 0.28 $\pm$ 0.1\% for groups 1 and 2, respectively.  The polarization angles ($\phi$) were -15 $\pm$ 6.6$^{\circ}$ and 4.8 $\pm$ 10.9$^{\circ}$ with respect to the optics plane.  Multiplication by the appropriate gain factors gives adjusted IP values of   $P$ = 0.54 $\pm$ 0.1\% for group 1 and $P$ = 0.53 $\pm$ 0.2\% for group 2.  

Oblique reflections induce polarization oriented perpendicularly to the plane of incidence \cite{renbarger98}.  For SMTO, this polarization should be dominated by the effects of the first three mirrors:  the tertiary and two periscope mirrors.  The latter have incident angles of 45$^{\circ}$, and will each induce horizontal polarization with respect to the optics plane.  Induced polarization from the tertiary will vary with the elevation angle of the object, but will have some component that will enter the VPM optics along the horizontal axis.  Thus, it seems reasonable that this should cause, at least in part, the horizontal IP that we have measured.

An IP of less than 1\% is comparable to that of many polarimeter systems currently or previously used, such as Hertz \cite{dowell98}.   In the future, we hope to characterize the nature of the Hertz/VPM IP  more thoroughly.

\section{Conclusions}\label{sec:conc}

We present the dual-VPM system as an alternative to the commonly used HWP.  This modulator, which operates in reflection with only small linear translations, is able to simulate the action of a rotating HWP.  The inherent properties of the dual-VPM modulator provide numerous benefits:  high durability, easy multiwavelength operation, and capability of observing circularly polarized light.  

Maximization of the dual-VPM system performance requires a full characterization of the performance of the grid-mirror system.  The VPM grid properties, in particular wire diameter and spacing, are important contributing factors to whether or not the grid will act ideally for a particular wavelength.  If wire grids are to be used at wavelengths shorter than those for which they are optimized, inteferograms can be used to determine proper spacing for half- and full-wave phase delays. 

Despite these difficulties, we found that the Hertz/dual-VPM polarimeter, as described above, operated with an instrumental polarization of less than 1\%, even with improper VPM and HWP settings.  Thus, the new device appears to be robust and is a competitive alternative for polarization modulation.  They are a viable option for mm/submm/IR astronomy applications, including the numerous experiments currently under development.  

\section*{Acknowledgments}

This material is based upon work supported by the National Aeronautics and Space Administration under Grant No. APRA04-0077-0150 issued through the Science Mission Directorate.  MK  was supported by GSRP Grant No. NNG05-GL31H.  HH was supported by a USRP grant at Goddard Space Flight Center.

% Place tables here

% Table 1 -- List of optics

\begin{sidewaystable}
\centering

\caption{Optics Elements of the Hertz/VPM Experiment and Their Properties} 
\label{tab:optics}

%\startdata
\begin{tabular}{cccc}

\textbf{Element} &\textbf{f/\# or focal length} & \textbf{Distance to next optic (mm)} & \textbf{notes} \\
\hline
telescope & 13.8 & 2410.6 & \\
periscope 1 & -- & 777& \\
periscope 2 & -- & 526 & \\
paraboloidal mirror 1 &  695\,mm & 620 & off-axis paraboloid\\
VPM1 & -- & 300 & grid 22.5$^{\circ}$ CCW from horizontal \\
VPM2 & -- & 225 & grid 45$^{\circ}$ CCW from horizontal \\
relay mirror 1 & -- & 537 & \\
paraboloidal mirror 2 & 695\,mm & 1161 & off-axis paraboloid \\
relay mirror 2 & -- & 225 & \\
relay mirror 3 & -- & 184 & \\
ellipsoidal mirror & 900\,mm (input) 360\,mm (output)  & 210 & Distance is to dewar window \\ 
Hertz & 3.9/3.5& -- & pupil lens/detectors\\
%\enddata
%\end{flushleft}
%\end{deluxetable}

\end{tabular}
\end{sidewaystable}
%\end{table}

\clearpage

%\rotate
\begin{sidewaystable}
\centering
\caption{Modulator Settings and Measured Efficiencies for Datafile Groupings}
\label{tab:settings}
%\startdata

%\hline
 \begin{tabular}{cccccccc}
  \hline
  \hline
&   &\raggedright{\mbox{               }VPM1 delay ($\mu$m)} & & \raggedright{\mbox{   }VPM2 delay ($\mu$m)} & & &\\
  Group & HWP angle ($^{\circ}$)& half & full & half & full & location &  efficiency\\
  \hline
 1 &  96 & 93 & 186 & 93 & 186  &SMTO & 54.3\% $\pm$ 2.6\% \\
 2 &  51 & 93 & 186 & 93 & 186  & SMTO & 53.8\% $\pm$ 1.2\% \\
3 &  68 & 100 & 193 & 82 & 175 &  lab & 47.7\% $\pm$ 15.6\% \\
 4 & 68 & 75 & 210 & 62.5 & 175  & lab & 52.3\% $\pm$ 1.8\%\\
 5 & 80 & 75 & 210 & 62.5 & 175 & lab  & 85.7\% +- 1.0\% \\
  \hline
\end{tabular}

 \end{sidewaystable}
 
 \clearpage

\begin{figure}[htbp]
   \centering
   \includegraphics[width=10.0cm]{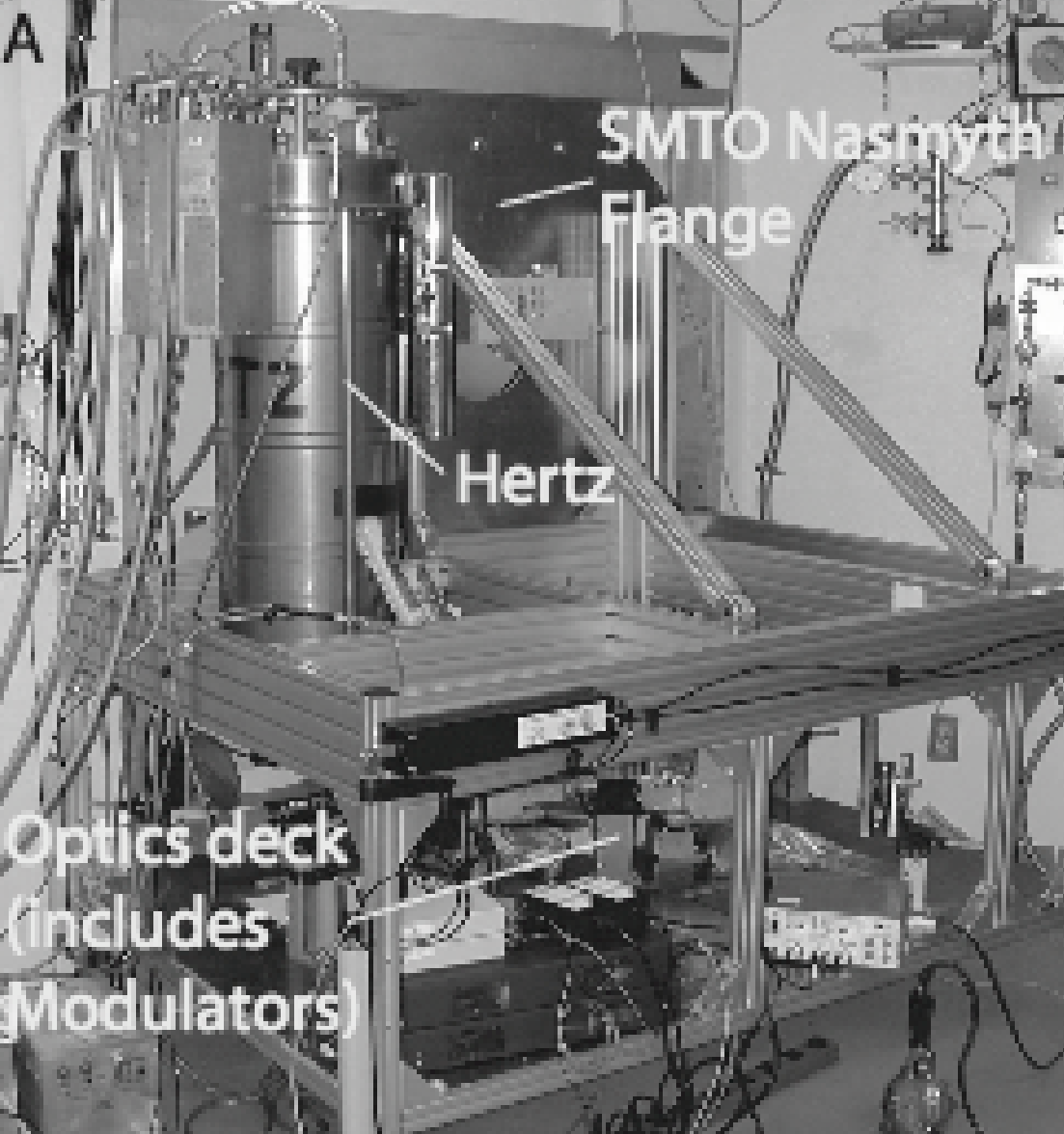}
   \caption{Photo of the Hertz/dual-VPM experiment at the SMTO.  Optics bench mounts to wall via a large flange; the optics sit at the Nasmyth focus. }
   \label{fig:mt}
\end{figure}

\begin{thebibliography}{99}

\bibitem{hiltner49} W.~A. Hiltner, ``Polarization of Light from Distant Stars by Interstellar Medium,''  Science, \textbf{109}, 165 (1949).

\bibitem{hall49} J.~S. Hall, ``Observations of the Polarized Light from Stars,'' Science \textbf{109}, 166 (1949).

\bibitem{chandraxx} S., Chandrasekhar, and E. Fermi, ``Magnetic Fields in Spiral Arms'', ApJ \textbf{118}, 113-115 (1953).

\bibitem{lazarian03} A. Lazarian, ``Physics of Grain Alignment,'' in \textit{ASP Conf. Ser. 215:  Cosmic Evolution and Galaxy Formation:  Structure, Interactions, and Feedback}, J. Franco, L. Terlevich, O, L—pez-Cruz, and I. Aretxaga, ed. (Astronomical Society of the Pacific, 2000), pp. 69-78.  

\bibitem{cudlip82} W.~I. Cudlip, I. Furniss, K.~J. King, and R.~E. Jennings, ``Far infrared polarimetry of W51A and M42,'' MNRAS \textbf{200}, 1169-1173 (1982).

\bibitem{hildebrand84} R.~H. Hildebrand, M. Dragovan, and G. Novak, ``Detection of submillimeter polarization in the Orion nebula,'' ApJ \textbf{284}, L51-L54 (1984).



\bibitem{jackson99} J.~D. Jackson, \textit{Classical Electrodynamics}, Third Edition (John Wiley \& Sons, 1999).

\bibitem{shinnaga99} H. Shinnaga, M. Tsuboi, and T. Kasuga, ``A millimeter polarimeter for the 45-m telescope at Nobeyama,'' PASJ \textbf{51}, 175-184 (1999).

\bibitem{siringo04} G. Siringo, E. Kreysa, L. A. Reichertz, and K. M. Menten,``A new polarimeter for (sub)millimeter bolometer arrays," Astron. Astrophys. 422, 751-760 (2004).

\bibitem{chuss06} D.~T. Chuss, E.~J. Wollack, S.~H. Moseley, and G. Novak, ``Interferometric polarization control,'' App. Opt. \textbf{45}, 5107-5117 (2006)

\bibitem{battistelli02} E.~S. Battistelli, M. DePetris, L. Lamagna, R. Maoli, F. Melchiorri, E. Palladino, and G. Savini, ``Far infrared polarimeter with very low instrumental polarization,'' 	arXiv:astro-ph/0209180v1.

\bibitem{hildebrand00} R.~H. Hildebrand, J.~A. Davidson, J.~L. Dotson, C.~D. Dowell, G. Novak, and J.~E. Vaillancourt, ``A Primer on Far-Infrared Polarimetry'', PASP \textbf{112}, 1215-1235 (2000).

\bibitem{platt91} S.~R. Platt, R.~H. Hildebrand, R.~J. Pernic, J.~A. Dotson, and G. Novak, ``100-$\mu$m Array Polarimetry from the Kuiper Airborne Observatory:  Instrumentation, Techniques, and First Results'', PASP \textbf{103}, 1193-1210 (1991).

\bibitem{voellmer06} G.~M. Voellmer, D.~T. Chuss, M. Jackson, M. Krejny, S.~H. Moseley, G. Novak, and E.~J. Wollack, ``A kinematic flexure-based mechanism for precise parallel motion for the Hertz variable-delay polarization modulator (VPM),'' in Proc. SPIE \textbf{6273}, 114 (2006).  

\bibitem{baars99} J.~W.~M. Baars, R.~N. Martin, J.~G. Mangum, J.~P. McMullin, and W.~L. Peters, ``The Heinrich Hertz Telescope and the Submillimeter Telescope Observatory'', PASP \textbf{111}, 627-646 (1999).

\bibitem{schleuning97} D.~A. Schleuning, C.~D. Dowell, R.~H. Hildebrand, S.~R. Platt, and G. Novak, ``HERTZ, A Submillimeter Polarimeter'', PASP \textbf{109}, 307-318 (1997).

\bibitem{novak97} G. Novak, J.~L. Dotson, C.~D. Dowell, P.~F. Goldsmith, R.~H. Hildebrand, S.~R. Platt, and D.~A. Schleuning, ``Polarized Far-Infrared Emission from the Core and Envelope of the Sagittarius B2 Molecular Cloud'', ApJ \textbf{487}, 320-327 (1997).

\bibitem{goldsmith90} P.~F. Goldsmith, D.~C. Lis, R. Hills, and J. Lasenby, ``High Angular Resolution Submillimeter Observations of Sagittarius B2'', ApJ \textbf{350}, 186-194 (1990).

\bibitem{dowell98} C.~D. Dowell, R.~H. Hildebrand, D.~A. Schleuning, J.~E. Vaillancourt, J.~L. Dotson, G. Novak, T. Renbarger, and M. Houde, ``Submillimeter array polarimetry with Hertz,'' ApJ \textbf{504}, 588-598 (1998).

\bibitem{houde01} M. Houde, R.~L. Akeson, J.~E. Carlstrom, J.~W. Lamb, D.~A. Schleuning, D.~P. Woody, ``Polarizing Grids, Their Assemblies, and Beams of Radiation'', PASP \textbf{113}, 622-638 (2001).


\bibitem{clemens90} D.~P. Clemens, B.~D. Kane, R.~W. Leach, and R. Barvainis, ``Millipol, a millimeter/submillimeter wavelength polarimeter - Instrument, operation, and calibration,'' PASP \textbf{102}, 1064-1076 (1990).

\bibitem{renbarger98} T. Renbarger, J. ~L. Dotson, and G. Novak, ``Measurements of submillimeter polarization induced by oblique reflection from aluminum alloy'', App. Opt \textbf{37}, 6643-6647 (1998).






\end{thebibliography}
\end{document}